\documentclass[baaa]{baaa}

\usepackage[pdftex]{hyperref}
\usepackage{subfigure}
\usepackage{natbib}
\usepackage{helvet,soul}
\usepackage[font=small]{caption}

\begin{document}


\journalvol{60}
\journalyear{2018}
\journaleditors{P. Benaglia, A.C. Rovero, R. Gamen \& M. Lares}


\contriblanguage{0}


\contribtype{1}

\thematicarea{5}

\title{Las antenas de espacio profundo en la Argentina}


\titlerunning{Las antenas de espacio profundo en la Argentina}


\author{M. Colazo\inst{1}}
\authorrunning{M. Colazo}


\contact{mcolazo@conae.gov.ar}

\institute{Comisión Nacional de Actividades Espaciales, Argentina 
}


\resumen{
Desde diciembre de 2012, fecha en que se inauguró en la provincia de Mendoza la antena de Espacio Profundo DS3 de la Agencia Espacial Europea, se concretó para nuestro país la posibilidad de uso de este equipamiento para las actividades espaciales y científicas. Se están llevando a cabo en este momento trabajos de desarrollo de aplicaciones con otras instituciones científicas del país para el uso astronómico de parte del tiempo de observación asignado a la Argentina. Algunos resultados ya han sido presentados en reuniones de la AAA. Ahora se suma a las capacidades argentinas la antena de Espacio Profundo que China ha instalado en la provincia de Neuquén como parte de su Programa de Exploración Lunar.  En este trabajo se describen las características de la nueva estación y el papel que tiene la Argentina en este proyecto a través de su agencia espacial, la CONAE.
}

\abstract{
Since December 2012, the Deep Space Antenna DS3 of the European Space Agency was inaugurated in the province of Mendoza. The possibility of using this equipment for space and scientific activities was promoted by our country. Several scientific institutions in the country are working together so that part of the observation time assigned to Argentina can be used by the astronomical community. Some results have already been presented at meetings of the AAA. Now the Deep Space antenna that China has installed in the province of Neuquén as part of its Lunar Exploration Program is added to the Argentine observational capabilities. This paper describes the characteristics of the new station and the role that Argentina has in this project through its space agency, CONAE.
}



\maketitle

\section{Introducción}
\label{S_intro}
La exploración y utilización del espacio ultraterrestre con fines pacíficos, desde las cercanías de la Tierra hasta el espacio profundo, constituye una herramienta importante de la inserción internacional del país en el ámbito de la investigación científico-tecnológica del espacio exterior así como en la participación en la instrumentación de misiones interplanetarias, en el marco de la cooperación internacional asociativa.

La Comisión Nacional de Actividades Espaciales de Argentina (\citeauthor{conae}) y la Agencia de Lanzamiento y Control de Satélites de China (CLTC) firmaron el 20 de julio de 2012 un acuerdo por el cual nuestro país dará apoyo mediante instalaciones de seguimiento terrestre, únicas en el hemisferio sur, a las misiones de exploración de la Luna programadas por China. Bajada del Agrio, en la provincia del Neuquén, fue el lugar elegido para construir la estación de seguimiento, comando y adquisición de datos de naves espaciales. Cuenta con una antena de 35 metros de diámetro que también se dedicará a la investigación del espacio profundo, es decir más allá del sistema Tierra – Luna. A través de la CONAE, la Argentina hará uso de estas instalaciones provistas con tecnología de avanzada, para el desarrollo de actividades de exploración interplanetaria e investigaciones en astronomía.

\section{El programa chino de exploración de la Luna y la red de estaciones para las misiones del espacio profundo}

El programa chino de exploración lunar es un programa de exploración espacial que incluye misiones robóticas y tripuladas a nuestro satélite natural. Este programa, anunciado en el año 2003 es liderado por la Administración Espacial Nacional de China, agencia responsable de la actividades espaciales chinas. 

\subsection{Estructura del programa espacial chino de exploración de la Luna}

El programa define tres hitos de exploración automática con un objetivo último, el alunizaje tripulado.

Fase I: Exploración desde la órbita lunar.
En esta primera fase las sondas Chang'{ e} 1 y Chang'{ e} 2 orbitaron la Luna \nocite{*}. La sonda Chang'{ e} (diosa china de la Luna) 1 fue lanzada el 24 de octubre de 2007 con el objetivo de realizar un mapeo tridimensional de la superficie de la Luna, reconocer las características del suelo lunar y explorar el entorno espacial entre la Luna y la Tierra. La misión duró 16 meses hasta que se estrelló en la superficie lunar. La sonda Chang'{ e} 2 fue lanzada el 1 de octubre de 2010 con el objetivo de recolectar imágenes de alta resolución de posibles sitios para el alunizaje en la siguiente fase del programa y también realizar varias pruebas tecnológicas cruciales para la continuidad del programa espacial.

Fase II: Alunizaje y reconocimiento robótico móvil.
El objetivo de esta fase es el alunizaje controlado y el despliegue de un rover lunar Yutu (Conejo de Jade) no tripulado para explorar la zona cercana al sitio de alunizaje. El rover fue diseñado para transmitir video en tiempo real, excavar y analizar muestras de polvo. Se presentaron problemas técnicos que limitaron la operacion en la misión que estaba calculada con una duración nominal de tres meses. La sonda utilizada fue la Chang'{ e} 3 que consistía de dos módulos, el vehículo alunizador y el vehículo explorador, el rover. Una segunda misión de alunizaje y exploración está planificada para fines de 2018 con el lanzamiento de la sonda Chang'{ e} 4, construida como respaldo para la misión Chang'{ e} 3.

Fase III: Extración robótica de muestras y recuperación.
La misión Chang'{ e} 5 está en desarrollo y se espera su lanzamiento para el año 2019. Será la primera misión china que traerá muestras del suelo lunar a la Tierra. 
El 23 de octubre de 2014 se lanzó la sonda Chang'{ e} 5-T1 que completó un recorrido alrededor de la Luna de 8 días e incluyó la evaluación de técnicas de reentrada a la atmósfera terrestre. Posteriormente, está planificado el lanzamiento de la sonda Chang'{ e} 6, y sus características quedarán definidas por los resultados obtenidos con la misión Chang'{ e} 5.

\subsection{La red de estaciones de espacio profundo}

La necesidad de dar soporte desde la Tierra a las misiones a la Luna impulsó la construcción de dos estaciones de Espacio Profundo en territorio chino. Una antena de 35 metros de diámetro se encuentra operando en Kashgar al noroeste de China. La segunda antena, de 64 metros de diámetro, fue instalada en Jiamusi, al noreste. El soporte de telemetría, seguimiento y comando (TT\&C) que pueden brindar estas dos estaciones deja una amplia zona fuera de cobertura. 

Con la instalación de una estación en sudamérica la cobertura para TT\&C llegaría a un 90\%. 

\section{La estación instalada en Neuquén}

\begin{figure}[!t]
  \centering
  \includegraphics[width=0.45\textwidth]{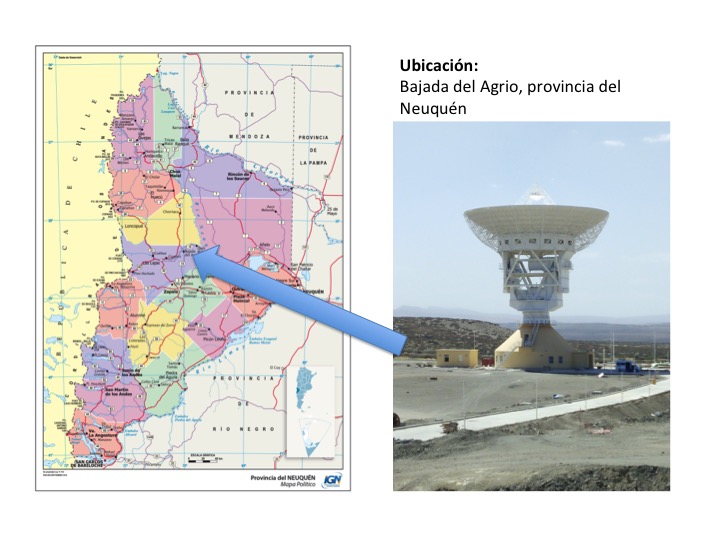}
  \caption{Ubicación de la antena de Espacio Profundo instalada en Argentina. Derecha: foto del instrumento.
}
  \label{ubicacion_antena}
\end{figure}

Después de un exhaustivo proceso de selección entre varios sitios candidatos, entre ellos países vecinos de Argentina, y gracias al apoyo de la Provincia de Neuquén y de la Comisión Nacional de Comunicaciones, se consideró que un lugar cercano a la localidad neuquina de Bajada del Agrio, cumple con todas las características necesarias para albergar la estación y la antena de exploración del espacio profundo  (Figura ~\ref{ubicacion_antena}).

Disponer de una estación de este tipo en nuestro territorio implica la oportunidad de participar en un importante programa de exploración de la Luna, y el acceso a nuevas herramientas y tecnologías para los proyectos espaciales de la CONAE. La agencia espacial argentina actuará como coordinadora entre CLTC y las autoridades nacionales, provinciales y locales que correspondan, a fin de brindar apoyo a las actividades necesarias para establecer y operar dicha estación en Neuquén.
Se trata de un proyecto de enorme relevancia para nuestro país, que permitirá desarrollar actividades de exploración interplanetaria, el estudio del espacio lejano, la observación astronómica, el seguimiento y control de satélites en órbita y la adquisición de datos científicos.

Como resultado de la participación en el establecimiento de la estación en nuestro país, la CONAE tendrá acceso a la utilización del 10\% del tiempo de uso operativo de la antena de espacio profundo, por año, no acumulable. Para esto, la CONAE podrá instalar, mantener y operar equipamiento de procesamiento de datos y coordinará con CLTC las actividades locales. De la misma manera que se hizo con la antena de Espacio Profundo de la Agencia Espacial Europea (ESA) instalada en Malargüe, Mendoza, se buscará el aprovechamiento científico del tiempo, como por ejemplo, la observación radioastronómica.  

\subsection{Características técnicas del instrumento}

\begin{itemize}
	\item Diámetro de la antena: 35 metros
	\item Tipo de montura: alt-azimutal
	\item Bandas de Operación:  Banda S (2 GHz), Banda X (8 GHz), Banda Ka (32 GHz)
	\item Tipo de polarización: circular (ambas polarizaciones al mismo tiempo)

	\item Ancho del haz a mitad de potencia (HPBW): 0.03$^{\circ}$ (Banda S), 0.01$^{\circ}$ (Banda X), 0.006$^{\circ}$ (Banda Ka) 
\end{itemize}

La estación entrará en operaciones a principios de 2018. Una vez operativa la estación se llamará a un concurso abierto para uso de parte del tiempo argentino en proyectos científicos.

\begin{acknowledgement}
Agradecemos a los Comités Organizadores Local y Científico de la última Reunión de la AAA por la posibilidad  de compartir con la comunidad científica el estado de las actividades de la CONAE en relación al Espacio Profundo.
\end{acknowledgement}


\bibliographystyle{baaa}
\small
\bibliography{biblio}
 
\end{document}